# Mobility−dependent Low−frequency Noise in Graphene Field Effect Transistors

Yan Zhang[*], E. E. Mendez[*,#], and Xu Du[*]

[*] Department of Physics and Astronomy, State University of New York at Stony Brook, Stony Brook, NY 11794-3800

[#] Center for Functional Nanomaterials, Brookhaven National Laboratory, Upton, NY 11973-5000

## Abstract

We have investigated the low-frequency 1/f noise of both suspended and on-substrate graphene field-effect transistors and its dependence on gate voltage, in the temperature range between 300K and 30K. We have found that the noise amplitude away from the Dirac point can be described by a generalized Hooge's relation in which the Hooge parameter $\gamma_H$ is not constant but decreases monotonically with the device's mobility, with a universal dependence that is sample and temperature independent. The value of $\gamma_H$ is also affected by the dynamics of disorder, which is not reflected in the DC transport characteristics and varies with sample and temperature. We attribute the diverse behavior of gate voltage dependence of the noise amplitude to the relative contributions from various scattering mechanisms, and to potential fluctuations near the Dirac point caused by charge carrier inhomogeneity. The higher carrier mobility of suspended graphene devices accounts for values of 1/f noise significantly lower than those observed in on-substrate graphene devices and most traditional electronic materials.

Graphene's exceptional properties create opportunities for a broad range of applications, among others, in electronics and sensors[1-8]. Inherent noise, especially low-frequency noise poses a practical limit on how small an input signal can be in broadband circuits. Understanding low-frequency noise in graphene devices is therefore a key step to increase the signal-to-noise ratio and improve the performance of circuits based on them.

Thus far, several groups have reported studies on the behavior of low-frequency noise in single-layer and few-layer graphene field-effect transistors (FETs)[9-16]. It has been observed that the low-frequency noise power in graphene FETs generally follows a 1/f frequency dependence. In aqueous solutios, it has been demonstrated that suspended graphene helps reducing the noise level[15]. However, the gate-voltage (or, equivalently, the charge carrier density) dependence has exhibited a variety of behaviors. In single-layer graphene nanoribbons with width of ~30nm, the low-frequency noise power density $S_V$ was found to follow Hooge's empirical relation[9, 17],

$$S_V(f) = \frac{\alpha_H V^{2+\beta}}{N f^\gamma} \equiv A \frac{V^{2+\beta}}{f^\gamma}, \qquad (1)$$

with $\beta \sim 0$, and $\gamma \sim 1$, and where $V$ and $N$ are the source-to-drain voltage and the total number of charge carriers in the conducting channel, respectively. $A$ is usually called the *noise amplitude* and $\alpha_H$ the *noise, or Hooge, parameter*. $A$ depends on the area of the sample, whereas $\alpha_H$ is an intensive parameter.

In devices with widths larger than 500nm, however, the gate-voltage dependence of noise did not show apparent agreement with Hooge's relation and the behavior was rather complicated. In the vicinity of the charge neutrality voltage (Dirac point), where the number of carriers is lowest, noise was at a minimum, contrary to Hooge's relation. With the gate voltage, $V_g$, increasingly away from the Dirac point, the noise increased until it reached a maximum at a sample-dependent gate voltage, beyond which the noise started to decrease. When both gate-voltage polarities are considered, graphically, the noise dependence on $V_g$ displayed an M-like shape. A simpler, V-shaped dependence, (which does not apparently follow Hooge's relation, either) has been found in bi-layer and multi-layer graphene samples.[10]

To account for these observations, several models have been proposed. For example, in a liquid-based field-effect device the observed M-shaped dependence was explained in terms of a charge noise model, in which, at low carrier density, noise was dominated by random charge fluctuations close to the graphene layer, and, at high density, by carrier scattering in the graphene layer[12]. The model also explained why the noise maxima occurred when the normalized transconductance, $(dI_d/dV_g)/I_d$, was the largest. On the other hand, in an in-vacuum FET device, the observed increase of noise with increasing carrier density was attributed to the decrease of minority charge carriers induced by charge impurities[13] (spatial-charge inhomogeneity model).

Furthermore, 1/f noise has been studied in a large number of devices exposed to air for extended periods of time (more than a month). The observed increase of noise with time was attributed to decreased mobility and increased contact resistance[16]. Overall, there has been no consensus on a unified relationship that can account for the diverse behavior of 1/f noise in graphene devices.

In this Letter, we report a comparative study of the low-frequency noise (henceforth denoted simply as noise) in suspended and on-substrate graphene devices. Our results are interpreted in terms of a modified Hooge model that also allows to explain results from previous reports. In our model, the Hooge parameter, $\alpha_H$, is not constant but variable, and affected by: a) scattering of the charge carriers, which contributes to a universal (i.e., sample independent) dependence of $\alpha_H$ on device mobility; and b) the dynamics of the scattering, such as trapping-detrapping, which, although makes a sample- and temperature-dependent contribution to $\alpha_H$ as a ($V_g$-independent) multiplication prefactor, does not manifest itself directly in the DC transport characteristics. Moreover, our model suggests that the gate-voltage dependence of noise is due to contributions from various scattering mechanisms, as well as charge carrier inhomogeneity near the neutrality point. It follows from the model that the Hooge parameter would be reduced by reducing charge trapping and the number of scattering centers[3,17], which is consistent with our observation of reduced noise in suspended graphene.

In our study, graphene flakes (with width ranging from 0.4μm-6μm) were mechanically exfoliated with Scotch tape from HOPG onto a 285nm-thick thermally grown $SiO_2$ film on top of a heavily p-doped Si substrate. Graphene FETs were fabricated with standard electron beam lithography and metallization (Cr/Au, 3/35 nm) methods. Prior to doing electrical measurements, the devices were annealed overnight in ultrapure oxygen at 180°C. Suspended graphene devices were prepared from conventional on-substrate devices and wet-etched with buffered oxide etch (BOE) 7:1 through a predefined PMMA window for five minutes for complete removal of the $SiO_2$ underneath graphene. Followed by BOE etching, the devices were transferred into hot acetone for removal of PMMA, and then into hot isopropanol. The samples were kept in liquid during processing and until they were finally taken out from hot isopropanol.[4]

Graphene devices were studied in a constant current source-bias configuration. DC voltage and voltage noise between source and drain of each device was amplified simultaneously with two separate SIM 910 voltage amplifiers with a 100 – gain factor. Noise spectra were recorded using a SR770 FFT spectrum analyzer with a 1000 – time linear average. The measurement setup was calibrated by measuring thermal noise from metal film resistors, and noise background was routinely checked to make sure it was well below the noise signal. All the measurements were carried out in vacuum, with a small amount of high-purity helium added as a heat-exchange gas for temperature-dependent measurements in the 30-300K range. In initial tests we did not observe any significant difference between two-terminal and four-terminal measurements, regarding contact resistance and noise, thus the measurements reported here were carried out in a two-terminal drain-source configuration.

In total, we measured the noise characteristics of six on-substrate, or non-suspended, graphene devices (NSG) and five suspended graphene devices (SG). In all cases, we found that the normalized voltage noise $S_V/V^2$ was independent of the external drain current throughout the temperature range of this work, indicating that the noise was due to resistance fluctuations[18].

The observed voltage noise generally followed the 1/f dependence of Hooge's relation[17] (Eq. 1) The noise power parameter, γ, was found to be $\gamma = 1.0 \pm 0.1$. The noise amplitude, A, was determined from a fit of the experimental $f*S_V(f)/V^2$ data to Eq. 1. Figures 1a and 1b show the gate-voltage dependence of the resistivity and the noise amplitude at room temperature for a typical on-substrate graphene device (NSG5) and a suspended graphene device (SG5), respectively. The charge carrier densities are proportional to $V_g$, with proportionality constants of $7.2\times10^{10}/cm^2$-V and $1.8\times10^{10}/cm^2$-V, for NSG5 and SG5, respectively. As seen in Figures 1a and 1b, in both samples the noise amplitude *A* increases monotonically with increasing carrier density. The V-like shape is contrary to Hooge's relation (Eq. 1) prediction, but is quite similar to the dependence found in bilayer and multi-layer graphene samples[10]. In other samples, such as NSG1 (see Fig. 2b), we observed an M-shaped dependence, with the noise amplitude increasing with increasing $V_g$ near the charge neutrality point but then decreasing at higher $V_g$, analogous to recent results in single-layer graphene and liquid-gated graphene transistors[12, 13].

Next we turn to the temperature dependence of the devices' resistance and noise characteristics. Figure 2a shows the resistivity versus gate voltage in an on-substrate device (NSG1) from T = 300K to 30K. The change of the resistance with temperature is quite small, especially away from the Dirac point. This is not surprising, since for graphene-on-$SiO_2$ devices the mobility is mainly governed by temperature-independent charged-impurity scattering[19]. In sharp contrast, the temperature dependence of noise is very strong, as seen in Figure 2b.(Note that NSG1 has an area ~ 35 times larger than NSG5.) The noise amplitude decreased monotonically with decreasing temperature, up to a factor of about 4 between 300K and 30K, a temperature range throughout which the noise spectrum was linear with 1/f dependence. Below 30K, a deviation from linearity was observed, perhaps due to the onset of random telegraph noise, which is outside the scope of this work. The very different dependence of the resistance and the noise on temperature highlights the sensitivity of the latter to microscopic processes to which the resistance is almost immune.

Similarly, we have studied the temperature dependence of the noise amplitude in suspended graphene devices. Figure 2c shows the resistivity versus gate voltage for device SG5 between 300K and 30K. As seen in the figure, the resistivity at the Dirac point increases much more with decreasing temperature than in the case of on-substrate graphene, as a result of a reduced residual carrier density[4, 20]. On the other hand, the mobility of SG5 (and other suspended devices) shows very weak temperature dependence, similar to on-substrate devices.

The dependence of the noise amplitude on gate voltage for SG5 at different temperatures is summarized in Fig. 2d, which is similar to that of other devices that exhibit a V-shape dependence. As temperature decreases from 300K to 30K, the noise amplitude decreases monotonically, except for an anomaly at T = 50K, at which the noise level is comparable to that between 105K and 145K. The overall noise reduction at 30K relative to 300K is about three times, that is, comparable to the reduction observed in NSG1 (see Figure. 2b). The anomalous behavior at T = 50K has been observed in other suspended devices, although at different temperatures. Its origin is not known.

To shed light on the results described above we have revisited Hooge's relation, which predicts that the noise amplitude should be inversely proportional to the total carrier number in the system: $A = \alpha_H / N$. We first consider the simplest case, where devices with different sizes but of similar quality and under identical gating and temperature conditions are compared for their noise amplitude. In this case, at any given gate voltage we expect the carrier number to be proportional to the channel area. As shown in Figure 3, we found that the noise amplitude of the devices indeed scales inversely with channel area, indicating that at least this aspect of Hooge's relation is satisfied: at each fixed gate voltage, the noise amplitude follows *1/N*. Therefore it seems justified to adopt Hooge's formal relation as the basis for understanding noise in our graphene devices.

The carrier number can also be changed by tuning the gate voltage. However, as mentioned before, the noise amplitude in graphene devices does not follow the simple $1/V_g$ dependence expected from Hooge's relation, but instead it shows a rather complicated behavior. This logically suggests that varying $V_g$ not only changes the carrier number, but also varies the properties of the channel itself and therefore the value of the Hooge parameter. Indeed, the value of $\alpha_H$ is not necessarily a constant, but it may instead depend on crystal quality and on the scattering mechanisms that determine the mobility $\mu$[21]. In a graphene device these include charged impurity scattering, short range disorder scattering, ripple scattering, etc.[22]. While the carrier mobility associated with charged impurity scattering has been shown to be $V_g$ independent, mobility associated with all the other scattering mechanisms does depend on $V_g$[22]. Thus it is reasonable to assume that $\alpha_H$ should also depend on $V_g$ rather than being constant.

Based on the above consideration, we characterize graphene by its mobility and look for a correlation between $\alpha_H$ and $\mu$. The calculated Drude mobility ($\mu=\sigma/ne$) and the Hooge parameter $\alpha_H = A \times N = (S_V/V^2) \times f \times N$ at T=300K are plotted in logarithmic scales on the upper part of Figure 4(a) for all the samples studied (both on-substrate and suspended graphene). Because of potential fluctuations induced by charged impurities, the carrier concentration in the vicinity of the Dirac point cannot be reduced to zero. Charge carrier density smaller than the residual charge density (typically $10^{11}$ cm$^{-2}$ or $V_g \sim$ few V in on-substrate devices, and $10^{10}$ cm$^{-2}$ or $V_g \sim$ 1V in suspended devices) is not considered for the calculations of $\alpha_H$ and $\mu$.

For each sample there are two curves (that in many cases practically overlap), corresponding to the electron and hole branches, such as those in Figures 1 (a) and (b). The correlation between $\alpha_H$ and $\mu$ is obvious: in all cases $\alpha_H$ decreases with increasing $\mu$. However, the slopes for suspended and on-substrate devices are quite different, being approximately − 1.5 and − 3, respectively, as shown in Figure 4(a). There is not a priori reason to believe that this significant difference is intrinsic; it may simply represent two different regimes of a common dependence. The fact that both devices with V-shape and M-shape noise characteristics have very similar $\Delta_H$ vs µ dependence strongly suggests that there is a physical phenomenon behind such a common dependence, and calls for further experimental and theoretical study.

In Figure 4(a), we have also included data extracted from the literature[9, 13, 16], plotted along our own data to test the generality of the $\Delta_H$ – µ. dependence described above. We note that the $\Delta_H$ vs. µ curve obtained from reference 8 has a singular slope of ~ -1, which leads to A= $\Delta_H$/N~1/µN~R, hence the maximum amplitude exhibited at the Dirac point in graphene nanoribbons[9]. Although this behavior appears qualitatively different from the results from all the other graphene devices, the physical models behind them are quite similar, with the only difference being the slope of the $\Delta_H$ vs. µ dependence. The deviation of the slope for graphene nanoribbons might be due to the change of the electronic structure in the geometrically confined devices.

The $\Delta_H$ - µ dependence shown in Figure 4(a) remains at lower temperatures, down to 30K, in both suspended and on-substrate graphene devices, although in general the lower temperature the smaller the value of $\Delta_H$. Since µ is almost independent of T [see the discussion above in relation to Figs. 2(a) and (c)] but $\Delta_H$ is not, we can factorize $\Delta_H$'s double dependence on µ and T:

$$\Delta_H \sim f(\mu) \times g(T), \qquad (2)$$

where *f(µ)* can be approximated as *(1/µ)$^\delta$* with $\delta$~1.5 and 3 for suspended and on-substrate devices, respectively, and *g(T)* is related to the (temperature-dependent) dynamic nature of the trapping-detrapping process, density fluctuations, etc. in the devices.

By treating g(T) as temperature and device dependent but mobility independent, we can get a "master" curve [shown in the lower part of Figure 4(a)] simply by dividing $\Delta_H$ in Figure 4(a) by different values (for different devices), so that all the curves now fall practically on top of each other. This behavior is consistent with our assumption that the "static" scattering makes a universal contribution to the noise amplitude from mobility fluctuations, while the actual values of the noise amplitude are also affected by "dynamic" contributions that do not have a direct correspondence in the DC transport characteristics.

In the following, we use the empirical relation between $\mu$ and $\alpha_H$ we have found to connect 1/$f$ noise to charge carrier scattering and to explain its dependence on gate voltage. We consider two $V_g$ regimes: one near the Dirac point (regime I), where local potential fluctuations give rise to electron-hole puddles and the effect of $V_g$ is mostly to modify the relative distribution of those puddles without changing much of the total carrier density; and another regime (regime II), far away from the Dirac point, where the potential fluctuations are relatively small compared to the gate voltage and the carrier density changes in proportion to $V_g$.

In graphene the mobility is limited by several scattering mechanisms. Here we focus on the two most important ones: short-range, disorder scattering and, long-range, Coulomb scattering (from charged impurities, etc.)[22]. Their different dependence on the carrier density leads to a contribution to the mobility that in the case of disorder scattering is inversely proportional to $V_g$ ($\mu_S = 1/C_S V_g$) whereas for Coulomb scattering is independent of it ($\mu_L = 1/C_L$).[23-25] $C_S$ and $C_L$ are short-range and long-range scattering constants, respectively, that depend on the density and strength of the corresponding scattering centers. Using Matthiesen's rule, $\mu = (1/\mu_S + 1/\mu_L)^{-1}$ and the empirical result $\alpha_H \sim 1/\mu$, we can write the noise amplitude as,

$$A = \alpha_H/N \sim (1/\mu)/N \sim (1/\mu_S + 1/\mu_L)/V_g \sim (C_S V_g + C_L)/V_g. \qquad (3)$$

The dependence of A on $V_g$ given by Eq. 3 is plotted in Figure 4(b) for a set of arbitrary $C_S$ and $C_L$ values. The solid lines in the figure show the separate contribution of the short-range and long-range terms as well as their combination. For small $V_g$ values, A decreases with increasing $V_g$ whereas for large enough $V_g$ A increases, regardless of the $C_S/C_L$ ratio. On the other hand, the crossover from one trend to another does depend on $C_S/C_L$, and the smaller the ratio, the larger the value of $V_g$ at which A starts to increase.

With these results at hand, we can now interpret the two very different behaviors observed in graphene devices in regime II ($V_g$ away from the Dirac point) described earlier: in some devices A keeps increasing with increasing $V_g$ up to the highest voltage we applied to them, while in many others the initial increase is followed by a definite decrease of A. The actual shape of the A vs $V_g$ curve depends on the interplay between the two major scattering mechanisms. In SG devices, for which Coulomb scattering is nearly absent (very small $C_L$), A is mainly affected by short-range disorder, which gives rise to A~ $V_g^{-1}$ from Eq. 3, or an increase of A with increasing $V_g$ for >1. For NSG, on the other hand, the Coulomb scattering dominates (large $C_L$), so that A ~ $1/V_g$, and A decreases with increasing $V_g$.

Now, the $1/V_g$ dependence, with its divergence as $V_g$ approaches zero, should dominate the region near the Diract point (regime I), regardless of the relative strength of the two scattering mechanisms, which is contrary to all our observations. To explain this discrepancy we need to keep in mind that near the Dirac point the graphene channel is not homogeneous but rather consists of local "patches" with different carrier densities. Such a complicated system can be seen as a network of resistors, each with a different number of carriers. In this regime, the effect of varying $V_g$ is to locally alter the imbalance between the number of electrons and holes, but the total number of carriers ($N_e + N_h$) remains approximately the same. A similar scenario has been proposed by Xu et al.[13] to explain the "dip" in the noise-$V_g$ dependence in terms of the sum of *normalized* current noise from conduction channels of electrons and holes in parallel. Here we generalize that idea, calculating the *total* noise (instead of the normalized noise) for resistors both in series and in parallel.

For simplicity, let us consider two resistors with resistances $R_1$ and $R_2$, and assume, for an easier discussion, that $R_i \sim 1/N_i$. This dependence implies that for low-carrier density Coulomb-scattering dominates, so that the conductance is proportional to the carrier density. If the resistors are in series, then the total "resistance" noise power density is given by

$$S_R = S_{R_1} + S_{R_2} = \frac{R_1^2 \alpha_H}{N_1} + \frac{R_2^2 \alpha_H}{N_2} \sim \frac{1}{N_1^3} + \frac{1}{(N-N_1)^3}. \tag{4}$$

In this case, the minimum noise should happen when $N_1 = N_2 = N/2$. It should be noted that such result can be generally reached as long as $R_i$ decreases monotonically with increasing $N_i$. If the resistors are in parallel, it is easier to express the noise in terms of the "conductance" noise,

$$S_\sigma = S_{\sigma_1} + S_{\sigma_2} = \frac{\sigma_1^2 \alpha_H}{N_1} + \frac{\sigma_2^2 \alpha_H}{N_2} \sim N_1 + N_2 = N, \tag{5}$$

which is independent of the imbalance between electrons and holes, as long as the total carrier number remains constant.

Generalizing these results to a combination of resistors in series and parallel, we infer that when we consider graphene as a network like that, then its total noise has its minimum when the channel is at its charge neutrality point, when overall all the patches have similar number of charge carriers ($N_1 = N_2$). As $V_g$ increases (but still within regime I) into the electron (or hole) branch, the number of carriers in the hole (or electron) puddles decrease, resulting in an increase in the total noise amplitude as long as these patches are partially in series with each other, which is a realistic assumption.

Once $V_g$ increases even further and the system is outside the potential-fluctuation regime, noise is determined by the nature of scattering, as explained earlier. For devices dominated by short-range scattering, noise keeps increasing monotonically with $V_g$, resulting in a V-shape profile. On the other hand, in devices with strong Coulomb scattering, noise decreases with increasing $V_g$, and a M-shaped profile is observed for the range of $V_g$ values used in our study.

In summary, we have studied low–frequency noise in suspended-graphene and graphene-on-$SiO_2$ devices. To explain the experimental data, we have used a generalized Hooge's relation in which the parameter $\alpha_H$ is not constant but decreases monotonically with the device's mobility. This model allows to correlate the noise amplitude A with the leading electronic scattering mechanisms, and explains well the diverse dependence of A on $V_g$ observed in a variety of graphene devices (including those in the literature) far from the Dirac point. On the other hand, that model fails to account for the observed increase of the noise amplitude with increasing $V_g$ near the Dirac point. This result is explained, though, in terms of a network of resistors in series and in parallel that mimic the charge imbalance and electron-hole puddles caused by potential fluctuations near the Dirac (charge-neutrality) point. As a result of the high carrier mobility, suspended graphene devices show low 1/f noise with $\alpha_H < 10^{-3}$ at room temperature, making them promising for low noise electronics and sensor applications.

**Acknowledgments.** This work has been funded by NSF under contract DMR-0705131 and by AFOSR under contract FA9550-10-1-0090. The authors thank L. Zhang for supplying HOPG and Si wafers, and B. Nielsen for technical support. This research was carried out in part at the Center for Functional Nanomaterials, Brookhaven National Laboratory, which is supported by the U.S. Department of Energy, Office of Basic Energy Sciences, under Contract No. DE-AC02-98CH10886.

Captions

Figure 1:

Resistivity and noise amplitude in single layer graphene device at room temperature as a function of gate voltage, $V_g$. (a) On-substrate device. Inset shows a typical optical image and the scale bar is 1μm. (b) Suspended device. Inset shows a typical SEM image and the scale bar is 2 μm.

Figure 2:

Temperature-dependent resistivity and noise amplitude in the temperature range of 30-300K. (a) Resistivity vs. gate voltage in on-substrate device at different temperatures. (b) Noise amplitude vs. gate voltage in on-substrate device at different temperatures. (c) Resistivity vs. gate voltage in suspended device at different temperatures. (d) Noise amplitude vs. gate voltage in suspended device at different temperatures.

Figure 3:

Noise amplitude vs. gate voltage for several NSG devices with similar quality but different sizes. The left panel shows the raw data, in which noise amplitude spans over 2 orders of magnitude for different samples. The right panel shows the area-scaled noise amplitude where all the curves roughly fall together. The insets show images of the devices.

Figure 4:

(a). Power law like dependence of the Hooge parameter on carrier mobility at room temperature. The hollow symbols correspond to on-substrate devices and the solid ones arefor suspended devices. The upper part shows the original values of $\alpha_H$, which are shown again in the lower part after having been divided by an arbitrary number to make them fall on the same master curve. The shifted data points are plotted using the same symbols as their original counterparts. The dashed line indicate the value of $\alpha_H=10^{-3}$, which usually is a lower limit to for conventional electronic materials. (b) Qualitative gate voltage dependence of noise amplitude for the Coulomb, short range, and mixed scattering. The dotted lines near the charge neutrality point represent the gate voltage dependence of the noise amplitude when the inhomogeneity of charge carriers is considered.

Figures

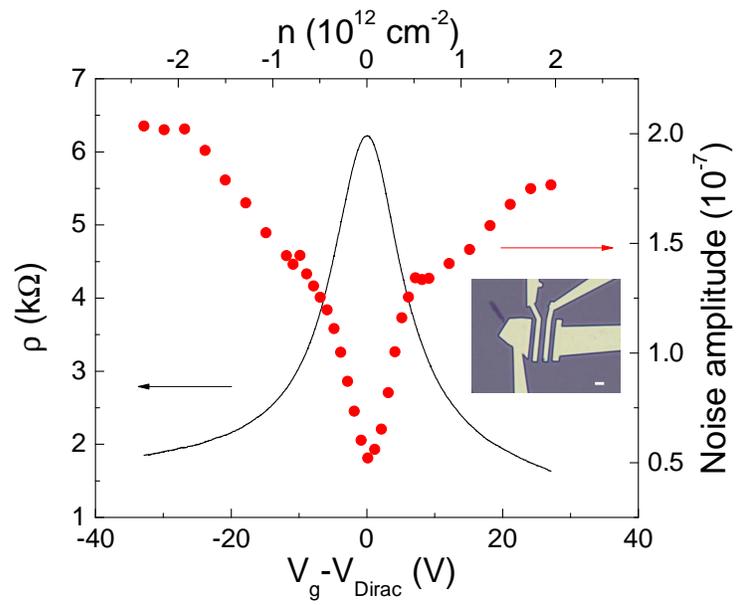

Figure 1(a)

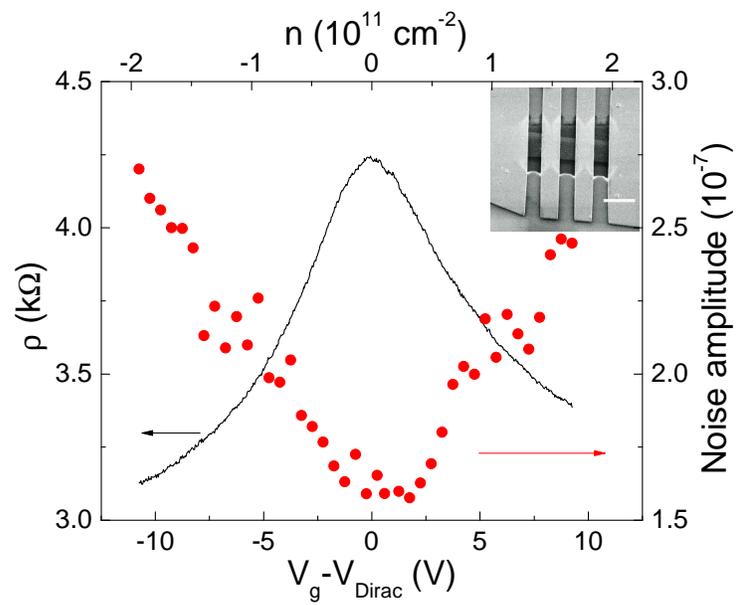

Figure 1(b)

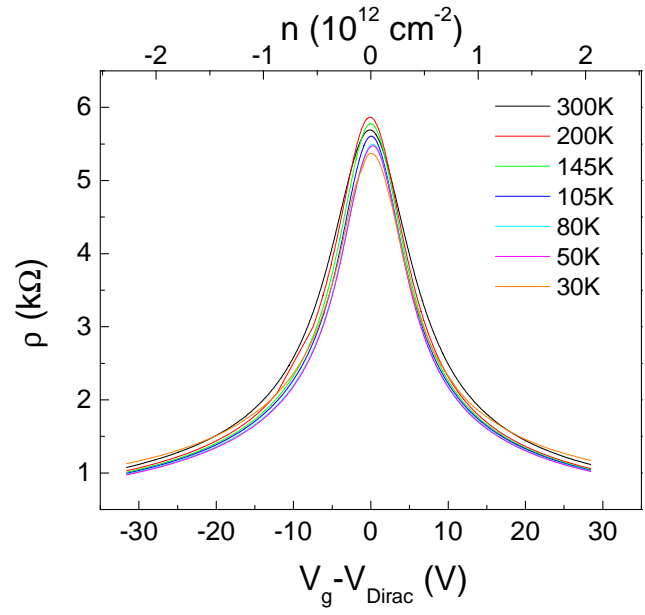

Figure 2(a)

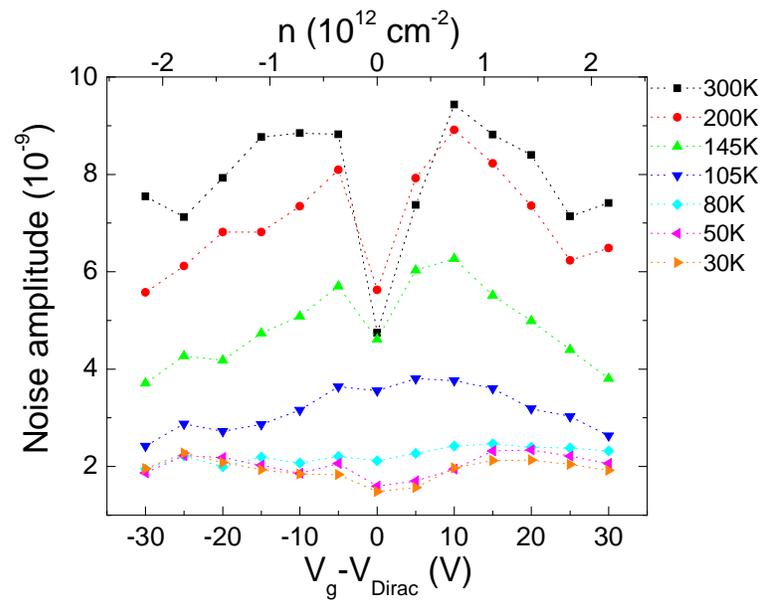

Figure 2(b)

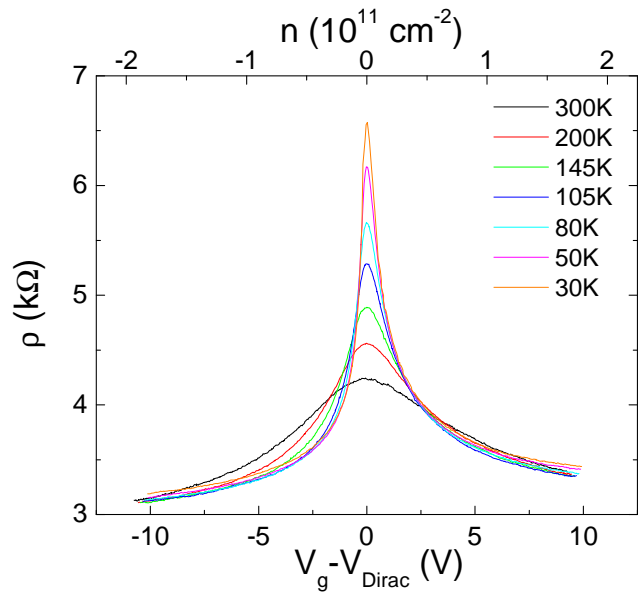

Figure 2(c)

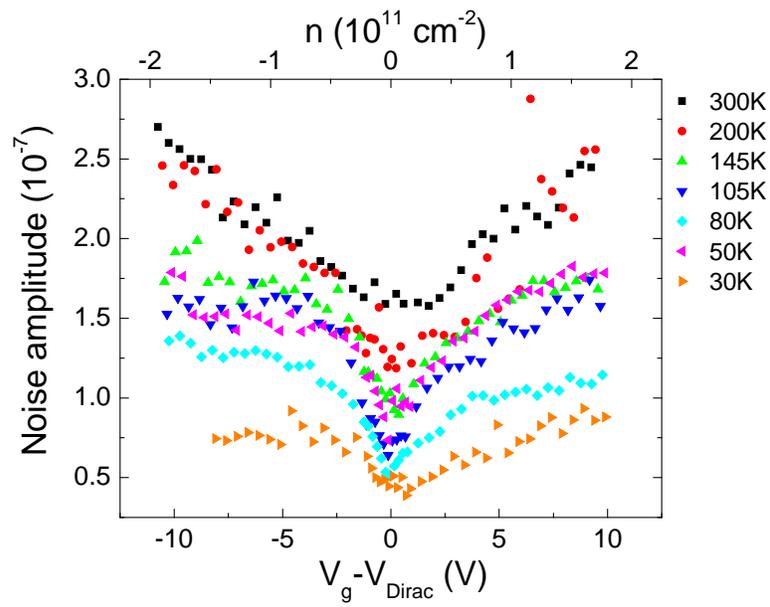

Figure 2(d)

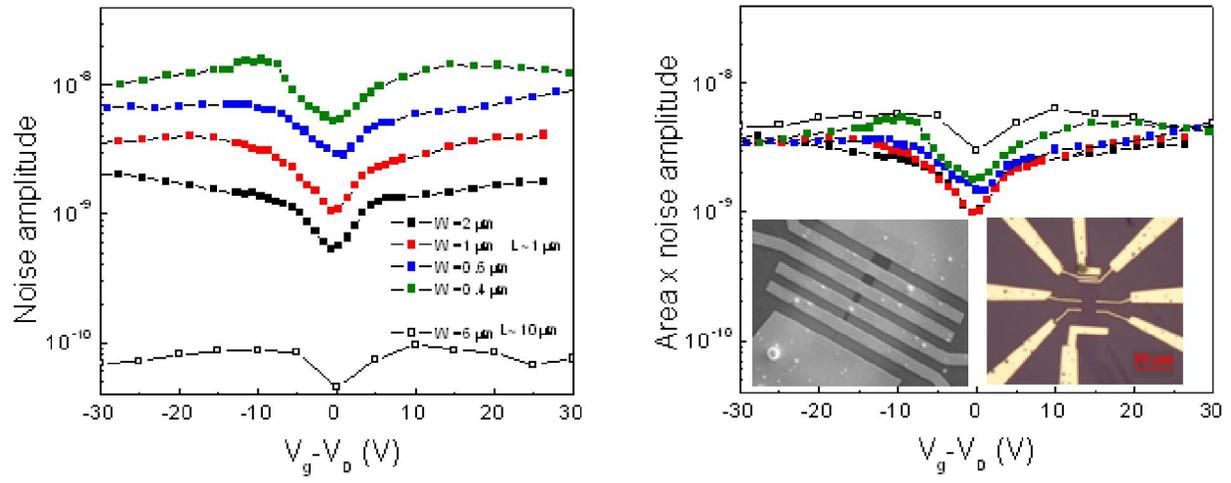

Figure 3

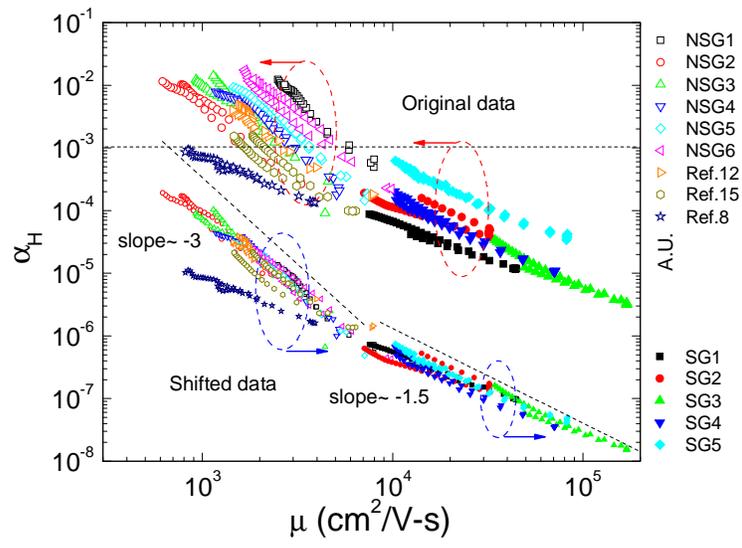

Figure 4(a)

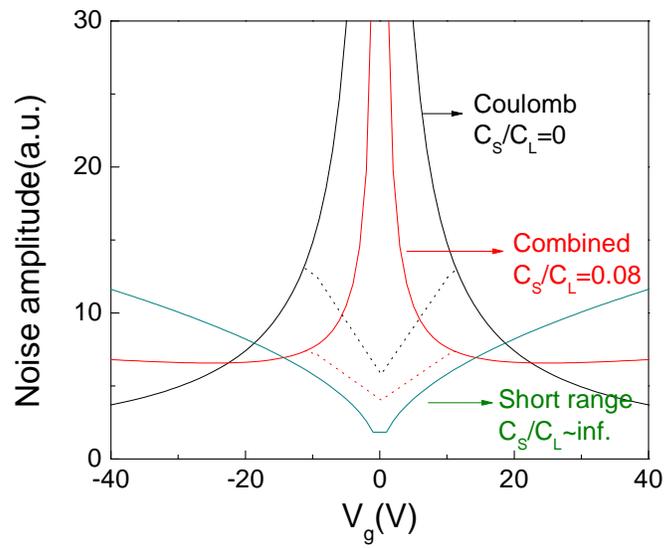

Figure 4 (b).